\documentclass[runningheads]{llncs}
\pdfoutput=1

\usepackage{graphicx}
\usepackage{amsmath}
\usepackage[bottom]{footmisc}
\usepackage{hyperref}

\begin{document}
\title{2.5D Thermometry Maps\\for MRI-guided Tumor Ablation}

\author{Julian Alpers\inst{1} \and
Daniel Reimert\inst{1,3} \and
Maximilian R\"otzer\inst{1} \and
Thomas Gerlach\inst{2} \and
Marcel Gutberlet\inst{3} \and
Frank Wacker\inst{3} \and
Bennet Hensen\inst{3} \and
Christian Hansen\inst{1}
}
%
\authorrunning{J. Alpers and D. Reimert et al.}
%
\institute{Faculty of Computer Science, University of Magdeburg, Magdeburg, Germany\\
\email{julian.alpers@ovgu.de}\\ \and
Faculty of Electrical Engineering and Information Technologies
Institute of Medical Technologies, University of Magdeburg, Magdeburg, Germany\\ \and
Institute for Diagnostic and Interventional Radiology, Medical School Hanover, Hanover, Germany}
%
\maketitle              
\begin{abstract}
Fast and reliable monitoring of volumetric heat distribution during MRI-guided tumor ablation is an urgent clinical need. In this work, we introduce a method for generating 2.5D thermometry maps from uniformly distributed 2D MRI phase images rotated around the applicator's main axis. The images can be fetched directly from the MR device, reducing the delay between image acquisition and visualization. For reconstruction, we use a weighted interpolation on a cylindric coordinate representation to calculate the heat value of voxels in a region of interest. A pilot study on 13 ex vivo bio protein phantoms with flexible tubes to simulate a heat sink effect was conducted to evaluate our method. After thermal ablation, we compared the measured coagulation zone extracted from the post-treatment MR data set with the output of the 2.5D thermometry map. The results show a mean Dice score of $0.75\pm0.07$, a sensitivity of $0.77\pm0.03$, and a reconstruction time within $18.02ms\pm5.91ms$. Future steps should address improving temporal resolution and accuracy, e.g., incorporating advanced bioheat transfer simulations.\makeatletter{\renewcommand*{\@makefnmark}{}
\footnotetext{The work of this paper is funded by the Federal Ministry of Education and Research within the Forschungscampus STIMULATE under grant numbers '13GW0473A' and '13GW0473B'. This work was also supported by PRACTIS - Clinician Scientist Program, funded by the German Research Foundation (DFG, ME 3696/3- 1).\\J. Alpers and D. Reimert - Joint first authorship\\B. Hensen and C. Hansen - Joint senior authorship}\makeatother}
\end{abstract}
\section{Introduction}
A wide range of minimally invasive therapies have been developed for cancer treatment, additionally to open surgery \cite{Ahmed.2014,mauri2017technical,tomasian2018percutaneous}. One of these methods is the use of microwave ablation (MWA). Especially for smaller tumors, MWA shows promising results for treatment \cite{tehrani2020use}. As the minimal ablative margin (MAM) is crucial for the local tumor progression (LTP), it is of greatest importance to assess if the malignancy has been adequately and completely treated, regardless of the etiology. For each millimeter increase of the MAM, a 30\% reduction of the relative risk for LTP was found. The MAM is especially important as the only significant independent predictor of LTP (p = 0.036) \cite{laimer2020minimal}. During the intervention, magnetic resonance (MR) imaging offers several advantages like a good soft-tissue contrast without the need of contrast agent, the free orientation and positioning of single slice scans and the possibility to accurately track changes in the temperature inside the tissue \cite{Gorny.2019,Kagebein.2018c,Rieke.2008,Senneville.2007}.\\
\\
\textbf{Contribution.} In this work, we propose a novel approach for the creation of a volumetric thermometry map without the development of a fully 3D sequence. The introduced 2.5D thermometry method utilizes any common 2D gradient-echo (GRE) sequences. Therefore, possible temporal limitations are less restricting than for the 3D sequences and images with higher resolution may be acquired offering standard thermometry accuracy of around $1^\circ$C deviation while being more robust towards MR inhomogeneities \cite{Gorny.2019}. We will show that our method is well-suited to reconstruct the actual coagulation zone after thermal ablation.\\
\\
\textbf{Related Work.} Zhang et al.\cite{zhang2019variable} propose a golden-angle‐ordered 3D stack‐of‐radial multi-echo spoiled gradient‐echo sequence with a variable flip angle. The image reconstruction is performed offline offering a temporal resolution between 2s-5s. Jiang et al.\cite{Jiang.2020} use an accelerated 3D echo-shifted sequence and the Gadgetron framework for image reconstruction. Temporal resolution lies at around 3s with a temperature error of less than $0.65^\circ$C. Quah et al.\cite{quah2020simultaneous} are aiming at an increased volume coverage for thermometry without multiple receive coils. An extended k‐space hybrid reconstruction was used, yielding an error of $<1^\circ$C and an acquisition time of 3.5s for each image. Fielden et al.\cite{Fielden.2018} present a comparison study between cartesian, spiral-out and retraced spiral-in/out (RIO) trajectories. Using the 3D RIO sequence, they achieved a true temporal resolution of 5.8s with a temporal standard deviation of $1.32^\circ$C. Marx et al.\cite{Marx.2014} introduced the MASTER sequence for volumetric MR thermometry acquisition, acquiring six slices in around 5s. In a later work \cite{Marx.2017} they use optimized multiple-echo spiral thermometry sequences, which yield a better precision than the usual 2D Fourier transform thermometry. Image acquisition takes between 7s-11s. Svedin et al.\cite{svedin2018multiecho} make use of a multi-echo pseudo-golden angle stack-of-stars sequence and offline image reconstruction using MATLAB. They achieved a temporal resolution of around 2s and a spatial average of the standard deviation through a time of $0.3-1.0^\circ$C. Odéen et al.\cite{odeen2016mr} propose the use of a 3D gradient recalled echo pulse sequence with segmented EPI readout. To estimate the temperature change, they also integrate a bioheat equation. They achieved a temperature root mean square error of $1.1^\circ$C. Golkar et al.\cite{golkar2018fast} introduce a fast GPU based simulation approach for cryoablation monitoring. The reconstruction takes 110s and the final result shows a Dice coefficient of 0.82. A summarize of the related work in comparison to our method is shown in Table \ref{tab:RelatedWork}.
\begin{table}[ht]
\centering
\caption{Overview about the related work in comparison to this work. Every work has been observed according to the following: 1) The kind of image sequence used. 2) The online or offline capability of the reconstruction framework. 3) The temporal resolution of the whole image acquisition in seconds. 4) The temperature accuracy in °C. 5) The resulting Dice Score similarity measurement if available.}
\begin{tabular}{|c|c|c|c|c|c|}
\hline
                       & \begin{tabular}[c]{@{}c@{}}Image \\ Sequence\end{tabular} & \begin{tabular}[c]{@{}c@{}}Reconstruction \\ Framework\end{tabular} & \begin{tabular}[c]{@{}c@{}}Temporal \\ Resolution {[}s{]}\end{tabular} & \begin{tabular}[c]{@{}c@{}}Temperature \\ Accuracy {[}°C{]}\end{tabular} & Dice Score \\ \hline
Zhang et al.{[}20{]}   & 3D                                                        & offline                                                             & 2 - 5                                                                  & ---                                                                      & ---                                                   \\ \hline
Jiang et al.{[}6{]}    & 3D                                                        & \begin{tabular}[c]{@{}c@{}}online \\ (Gadgetron)\end{tabular}       & 3                                                                      & \textless 0.65                                                           & ---                                                   \\ \hline
Quah et al.{[}13{]}    & Stack of 2D                                               & hybrid                                                              & 3.5                                                                    & \textless 1                                                              & ---                                                   \\ \hline
Fielden et al.{[}3{]}  & 3D                                                        & online                                                                 & 5.8                                                              & 1.32                                                                     & ---                                                   \\ \hline
Marx et al.{[}10{]}    & Stack of 2D                                               & ---                                                                 & 5                                                                      & 1.3                                                                      & ---                                                   \\ \hline
Marx et al.{[}9{]}     & Stack of 2D                                               & online                                                                 & 7 - 11                                                                 & 0.29 - 0.65                                                                      & ---                                                   \\ \hline
Svedin et al. {[}17{]} & 3D                                                        & \begin{tabular}[c]{@{}c@{}}offline \\ (MATLAB)\end{tabular}         & 2                                                                      & 0.3 - 1.0                                                                & ---                                                   \\ \hline
Odeen et al.{[}12{]}  & Stack of 2D                                               & \begin{tabular}[c]{@{}c@{}}offline \\ (MATLAB)\end{tabular}                                                                 & 2.4 - 4.8                                                                    & \textless 1.1                                                                      & ---                                                   \\ \hline
Golkar et al.{[}4{]}   & 3D                                                        & ---                                                                 & 110                                                                    & ---                                                                      & 0.82                                                  \\ \hline
This work             & Single 2D                                               & online                                                              & variable                                                               & 1                                                                        & 0.75                                                  \\ \hline
\end{tabular}
\label{tab:RelatedWork}
\end{table}
\section{Material and Method}
\label{sec:MaterialAndMethod}
\subsection{Image Acquisition}
\label{subsec:ImageAcquisition}
The proposed 2.5D thermometry relies on sampling the volume of interest (VOI) using a common 2D GRE sequence and [1,...,n] different orientations. The GRE sequence can directly reconstruct magnitude and phase images simultaneously. To ensure a proper sampling of the VOI in our setup the GRE sequence is rotated by $22.5^\circ$ around the applicator's main axis. This results in an evenly distributed sample of eight different orientations. To increase the spatial resolution, the angles between the acquired scans should be as high as possible, resulting in the following acquisition order: $0 ^\circ$, $90 ^\circ$, $45 ^\circ$, $135 ^\circ$, $22.5 ^\circ$, $112.5 ^\circ$, $67.5 ^\circ$, and $157.5 ^\circ$. To reduce the delay between image acquisition and visualization of the volumetric thermometry map, the SIEMENS Healthineers Access-I Framework was integrated. The framework allows for fetching the image data directly from SIEMENS MR devices without an intermediate imaging archive system.
\subsection{2.5D Thermometry Reconstruction}
\label{subsec:25DThermometryReconstruction}
\begin{figure}[ht]
    \centering
    \includegraphics[width=1\textwidth]{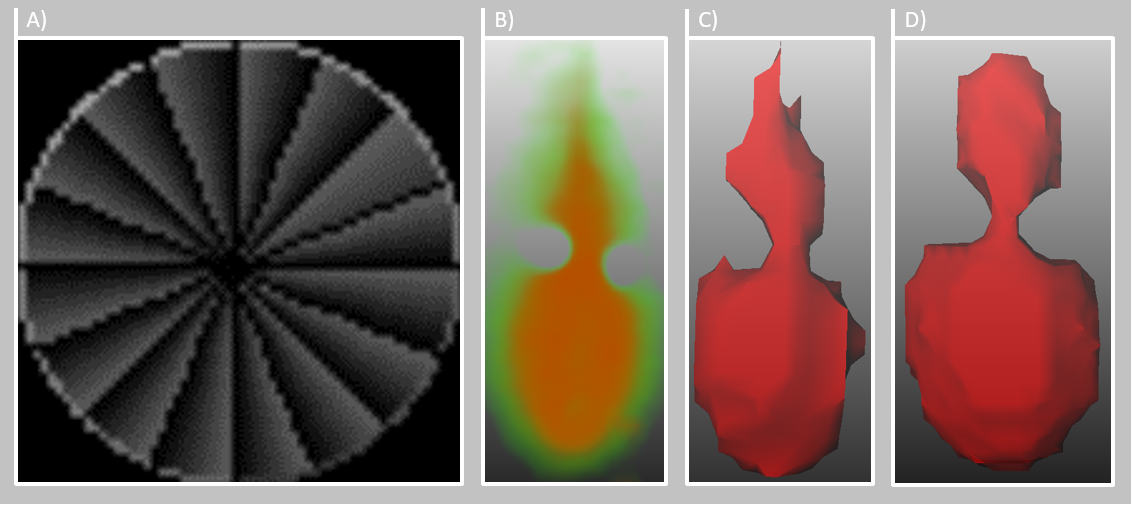}
    \caption{A) Example population map for output weights color coded on gray scale. B) Reconstructed volumetric heat map. C) Estimated coagulation necrosis based on a threshold of $57^\circ$C. D) Manually segmented ground truth.}
    \label{fig:Reconstruction}
\end{figure}

Before treatment starts, reference phase images are acquired for each of the eight orientations. Each newly acquired phase image will start computing the up-to-date 2D thermometry map for the current orientation during the treatment. To do so, the proton resonance frequency shift (PRFS) method is used as described by Rieke et al. \cite{Rieke.2008}. The temperature $T$ based on the PRFS is computed using the following Equation
\begin{equation}
\label{eq:PRFS}
    T = \frac{\phi(t) - \phi (t_0)}{\gamma \alpha B_0 TE} + T_0
\end{equation}
with $\phi(t) - \phi (t_0)$ defining the phase difference between the current time point $\phi(i)$ and the reference timepoint $\phi(i_0)$, $\gamma = 42,576\frac{MHz}{T}$ representing the gyromagnetic ratio of hydrogen protons, $\alpha = 0.01\frac{ppm}{\Delta T}$ representing the proton resonance frequency change coefficient, $B_0$ representing the used magnetic field strength and $TE$ representing the used echo time. The constant $T_0$ needs to be added to the temperature since Equation \ref{eq:PRFS} otherwise only computes the temperature change, neglecting the tissue's base temperature. The Access-I integration and 2D thermometry computation were implemented as modules using MeVisLab 3.4.1 \cite{ritter2011medical}. 
\begin{figure}[ht]
    \centering
    \includegraphics[width=1\textwidth]{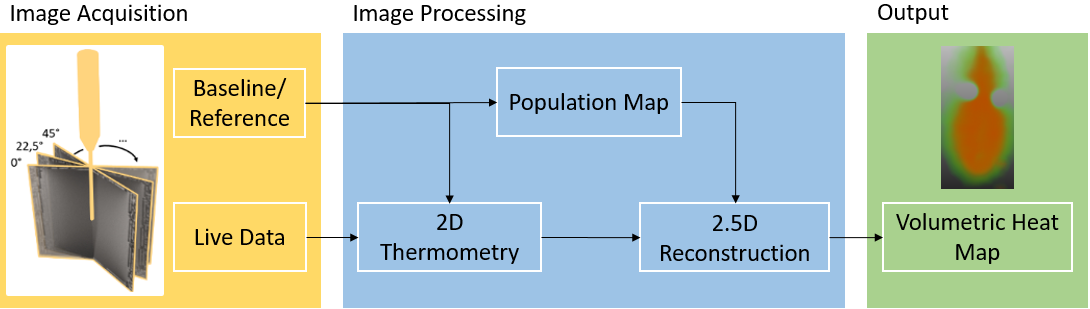}
    \caption{Schematic overview of the proposed method.}
    \label{fig:MethodOverview}
\end{figure}
The 2.5D thermometry reconstruction itself was implemented using C++. A schematic overview of the method can be seen in Figure \ref{fig:MethodOverview}. To handle the voxel values during slice rotation every cartesian coordinate was mapped to the corresponding cylindrical coordinate representation using Equation \ref{eq:CartesianToPolar},
\begin{align}
\label{eq:CartesianToPolar}
    P_r(x,y,z) &= P_c(r, \theta, z)\\
    r &= \sqrt{\left(x-x_c \right)^2 + \left(y-y_c\right)^2}\nonumber \\
    \theta &= atan2\left(\frac{x-x_c}{y-y_c}\right)\nonumber
\end{align}
where $x,y$ represents the Cartesian coordinates of the current voxel and $x_c,y_c$ represents the Cartesian coordinates of the centerline corresponding to the applicator's axis for every slice $z$ in the reconstructed volume. Upon acquisition of the reference images, a multi-dimensional population map is created. For each voxel $(x_i,y_i,z_i)$ in the reconstructed volume, this population map holds information about the radius $r$ and angle $\theta$ of the cylindrical coordinates, the general interpolation weight $I_w$, the adjacent interpolation partner coordinates $IP_{left}(x,y)$ and $IP_{right}(x,y)$ in the 2D live data as Cartesian representation and the weights $w_1$ and $w_2$ of those interpolation partners. The weights may be acquired using Equation \ref{eq:Weights},
\begin{align}
\label{eq:Weights}
    w_{1} &= \left\vert\frac{\theta_{IP_{left}} - \theta_{i}}{\theta_{IP_{left}} - \theta_{IP_{right}}}\right\vert\\
    w_2 &= 1 - w_1\nonumber
\end{align}
with $\theta_{i}$ representing the cylindric angle of the current Voxel $i$ and $\theta_{IP_{left}}$, $\theta_{IP_{right}}$ representing the orientation angles of the left and right interpolation partners, respectively. The 2D population map can be applied to every slice of the final 3D output volume, reducing the computational power needed. During the intervention, every acquired live image triggers the reconstruction of the up-to-date 2.5D thermometry map. Here, the heat value for each voxel is reconstructed using Equation \ref{eq:Reconstruction},
\begin{equation}
\label{eq:Reconstruction}
    T_i = I_w \cdot (w_{1} \cdot T_{IP_{left}} + w_{2} \cdot T_{IP_{right}})
\end{equation}
with $T_i$ representing the temperature of the current voxel $i$ and $T_{IP_{left}}$, $T_{IP_{right}}$ representing the temperature of the adjacent interpolation partners. Occurring vessels or other structures, which cause a heat sink effect are segmented during the intervention planning. Subsequently, the segmented structure is saved as an additional Look-Up Volume. Here, each voxel can be checked if it is part of a heat sink structure. Using this knowledge, the interpolation weight $I_w$, which ranges between $[0,1]$, may be adjusted. Figure \ref{fig:Reconstruction} shows a single dimension of the population map for parameter weighting, a reconstructed heat map, a coagulation estimation based on an empirically defined threshold and the corresponding ground truth segmentation. The source code is available for download at \url{https://github.com/jalpers/2.5DThermometryReconstruction}.
\subsection{Evaluation}
\textbf{Phantom design.} To create a first proof of concept, a pilot study was conducted to evaluate the 2.5D thermometry reconstruction using 13 bio protein phantoms as described by Bu Lin et al.\cite{BuLin.2008}. The coagulation zone's visibility in the post-treatment data sets increased by adding a contrast agent ($0,5\mu mol/L$ Dotarem) to the phantoms. For six phantoms, additional polyvinyl chloride (PVC) tubes with a diameter of $5mm$ and a wall thickness of $1mm$ were integrated into the phantoms (three single-tubes, three double-tubes) to simulate a possible heat sink (HS) effect.\\
\\
\textbf{Experimental setup.} The applicator of the permittivity feedback control MWA system (MedWaves Avecure, Medwaves, San Diego, CA, USA, 14G) was placed inside the phantom by sight and secured in position. Subsequently, the phantoms were placed inside a $1,5T$ MR scanner (Siemens Avanto, Siemens Healthineers, Germany). The coaxial cables connected to the applicator and MW generator were led through a waveguide. Chokes and electrical grounding measures were added as described by Gorny et al.\cite{Gorny.2019} to reduce radio frequency interference. In the case of the perfusion phantoms, the PVC tubes were led through the wave guide. They were connected to a diaphragm pump and a water reservoir outside the scanning room. A flow meter (SM6000, ifm electronic, Essen, Germany) was interposed between the reservoir and the pump, providing a flow rate of $800mL/min$. Observations showed a moderate HS effect using this setup with a maximum antenna power of 36W. Additionally, temperature sensors were inserted in two phantoms to experimentally verify the temperature accuracy of $1 ^\circ C$.
\begin{figure}[ht]
    \centering
    \includegraphics[width=1\textwidth]{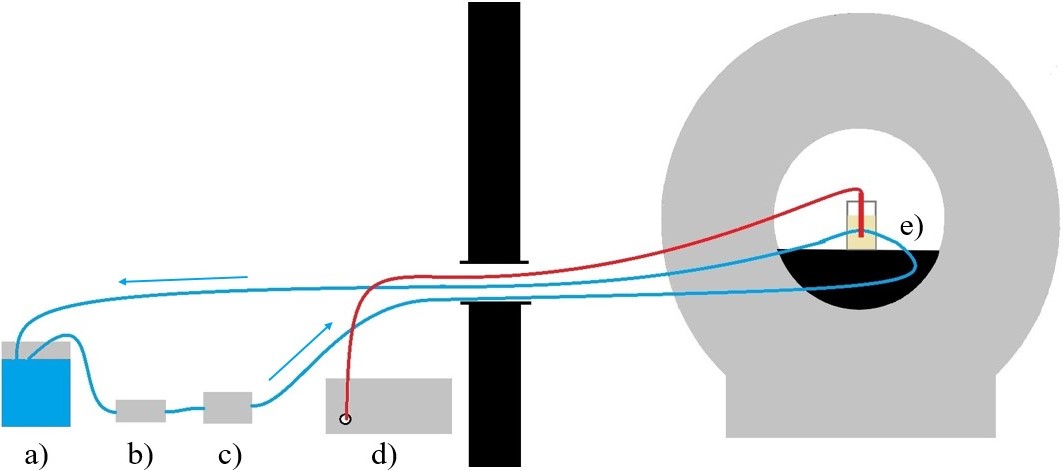}
    \caption{Experimental evaluation setup. Flexible tubes (blue) lead the water (a) through a flow meter (b), a diaphragm pump (c) and the bio protein phantom (e). The coaxial cables (red) connect the applicator with the MW generator d).}
    \label{fig:StudyDesignSketch}
\end{figure}
Right before treatment, ten reference phase images were acquired and averaged for each orientation to compensate for static noise. The MWA duration was set to 15 minutes with a temperature limit of $90^\circ$C. The GRE sequence offers a slice thickness of $5mm$, a field of view (FOV) of $256mm * 256mm$, a matrix of $256 * 256$, and a bandwidth of $260 Hz/Px$. Image acquisition took around $1.1s$ with a $5s$ break to simulate the temporal resolution for a breathing patient. The TE was $3.69ms$, the TR $7.5ms$, and the flip angle $7 ^\circ$. For post-treatment observation a 3D turbo spin echo (TSE) sequence (TE = $156ms$, TR = $11780ms$, flip angle = $180 ^\circ$, matrix = $256 * 256$, FOV = $256mm * 256mm$, bandwidth = $40 Hz/Px$, slice thickness = $1mm$) was used. The 3D TSE allows for proper visualization of the real coagulation zone due to a very high tissue contrast. Extraction of the coagulation ground truth was done manually by a clinical expert using MEVIS draw (Fraunhofer MEVIS, Bremen, Germany). All data sets used are available for download at \url{http://open-science.ub.ovgu.de/xmlui/handle/684882692/89}.\\
\\
\textbf{Statistical evaluation.} Final evaluation of the acquired data was performed using the dice similarity coefficient (DSC) as explained in Equation \ref{eq:DSC}
\begin{equation}
    \label{eq:DSC}
    DSC = \frac{2*TP}{2*TP + FP + FN}
\end{equation}
with $TP$ representing the true positives, $FP$ the false positives and $FN$ the false negatives. Additionally, the standard error of the mean (SEM) was computed at a confidence level of 95\% (p = 0.05) using Equation \ref{eq:SEM}
\begin{align}
    \label{eq:SEM}
    \sigma &= \sqrt{\frac{\sum(x_i-\bar{x})^2}{N-1}}\\
    SEM &= \frac{\sigma}{\sqrt{N}} * 1.96\nonumber
\end{align}
with $\sigma$ representing the standard deviation, $x_i$ the current sample, $\bar{x}$ the mean value and $N$ the sample size. To compute the SEM at a confidence level of 95\% it has to be multiplied by 1.96, which is the approximated value of the 97.5 percentile of the standard normal distribution.
\section{Results}
\begin{figure}[ht]
    \centering
    \includegraphics[width=1\textwidth]{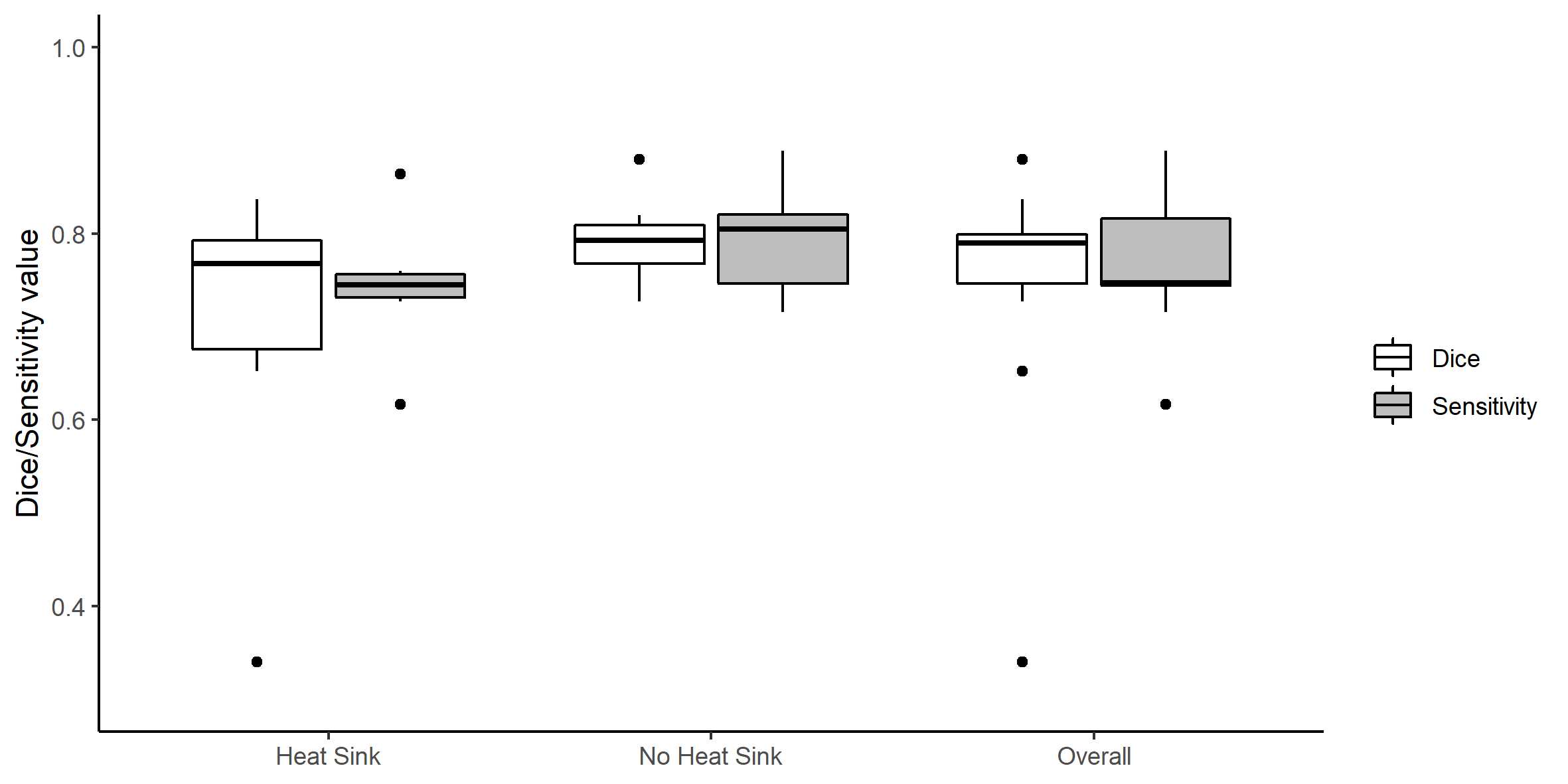}
    \caption{Summarized evaluation results for phantoms without HS effect, phantoms with HS effect and the overall results. Note that the data range [0,0.3] was left out because no data points are present in that range.}
    \label{fig:SummarizedResults}
\end{figure}

Summarized evaluation results can be seen in Figure \ref{fig:SummarizedResults}. Empirically determined coagulation thresholds were set between $51^\circ$C and $61^\circ$C depending on each phantom's pH value.
It is noticeable that the DSCs for HS phantoms show a very high SEM with $0.70\pm0.15 (\pm21.25\%)$ and $0.74\pm0.06 (\pm8.49\%)$ regarding the sensitivity. The high range results from a corrupted dataset due to heavy artifacts within the image data. Leaving the corrupted dataset out of the evaluation, the SEM shows a significantly lower deviation of $0.76\pm0.062 (\pm8.07\%)$ and $0.77\pm0.048 (\pm6.25\%)$ for the DSC and sensitivity, respectively. Observations show a slightly higher DSC and sensitivity for phantoms without any HS effect. Here, the values range from $0.79\pm0.04 (\pm 4.53\%)$ and $0.79\pm0.04 (\pm 5.55\%)$, respectively. Evaluation showed an overall SEM for the DSC of $0.75\pm0.07 (\pm9.76\%)$ and a SEM for sensitivity of $0.77\pm0.04 (\pm4.99\%)$. To evaluate the computational effort, every major step was performed 100 times. The creation of the population map and the heat sink look up volume took $25.53ms\pm 3.33ms$ and $3.91s\pm 0.59s$, respectively. These two steps need to be done just once before start of the treatment. The reconstruction of the 2.5D thermometry map was performed in $18.02ms\pm 5.91ms$ on a customary workstation (Intel( R) Core(TM) i5-6200U CPU, double-core 2.30GHz, 8GB RAM, Intel(R) HD Graphics 520). This reconstruction will be performed every time a new image is acquired during treatment.

\section{Discussion and Conclusion}
The aim of this work is to develop a volumetric thermometry map, which can be applied to a wide variety of clinical setups. Therefore, our work heavily relies on the up-to-date standard 2D GRE sequence for image acquisition. This allows for the standard accuracy of the thermometry up to $1.0^\circ$C. Nonetheless, the sampling of the 3D volume also results in some disadvantages, which need to be addressed in the future. First, the diffusion of the heat inside the tissue is not linear over time. Therefore, it would be necessary to include an adaptive temporal and spatial resolution depending on the current intervention time. A new study should be conducted to identify an optimal sequence protocol for this 2.5D thermometry approach. Second, we found that the reconstruction sometimes shows stair-case artifacts. Because only one image is acquired every few seconds, the time difference between adjacent orientations may be very high. The temperature difference for each voxel dependent on the applicator's radius may be computed and applied to the corresponding voxel on every other out-of-date data to compensate for this error. This transfer of the heat gradient may improve the reconstruction accuracy. Another approach may be the use of a model-based reconstruction to take different tissue characteristics into account. To pseudo-increase the temporal resolution, bio heat transfer simulations may also be included during reconstruction. The acquired live data may be able to adjust the simulation parameters to increase the simulation accuracy. Finally, our study only performs on bio protein phantoms. Results show a proof of concept for the proposed method, but it still has to be evaluated in real tissue and a more realistic clinical environment. Therefore, perfused ex vivo livers may be a way to go in the future. Additionally, we currently assume a breath-holding state or at least a breath-triggered image acquisition. Research shows that a wide range of interventional registration methods is available, but further investigations in this area still need to be done to create an applicable method. The last issues arise because of the MR inhomogeneity during image acquisition. The slightest disturbances may result in heavy image artifacts. Proper shielding of the MW generator is needed to reduce the SNR loss over time thus increasing the thermometry and reconstruction accuracy. \\
\\
In conclusion, we proposed a novel method for 2.5D thermometry map reconstruction based on common GRE sequences rotated around the applicator's main axis. A pilot study was conducted using bio protein phantoms to simulate cases with possible heat sink effects and without. The evaluation shows promising results regarding the DSC of the reconstructed 2.5D thermometry map and a manually defined ground truth. Future work should address the reconstruction method's improvement by integrating further apriori knowledge like the estimated shape of the heat distribution. Furthermore, a more realistic study should be conducted with bigger sample size and real tissue. In sum, the method shows a high potential to improve the clinical success rate of minimally invasive ablation procedures without necessarily hampering the standard clinical workflow of the individual clinician.
%

\end{document}